\documentclass[aps,prl,twocolumn,showpacs,superscriptaddress,groupedaddress]{revtex4}

\usepackage[utf8x]{inputenc}
\usepackage{amsmath,amssymb,amsfonts,subfigure,braket,color,dsfont}
\usepackage[english]{babel}
\usepackage{graphicx} 
\usepackage{sistyle}
\usepackage{listings}

\begin{document}

\title{Dispersive coupling between light and a rare-earth ion doped mechanical resonator}
\author{Klaus M{\o}lmer} 
\affiliation{Department of Physics and Astronomy, Aarhus University, Ny Munkegade 120, DK-8000 Aarhus C, Denmark.}
\author{Yann Le Coq}
\affiliation{LNE-SYRTE, Observatoire de Paris, PSL Research University, CNRS, Sorbonne Universit\' es, UPMC Univ. Paris 06, 61 avenue de l'Observatoire, 75014 Paris, France}
\author{Signe Seidelin}
\affiliation{Univ. Grenoble Alpes, CNRS, Inst. NEEL, F-38000 Grenoble, France}
\affiliation{Institut Universitaire de France, 103 Boulevard Saint-Michel, F-75005 Paris, France}

\date{\today}

\begin{abstract}

By spectrally hole burning an inhomogeneously broadened ensemble of ions while applying a controlled perturbation, one can obtain spectral holes that are functionalized for maximum sensitivity to different perturbations. We propose to use such hole burnt structures for the dispersive optical interaction with rare-earth ion dopants whose frequencies are sensitive to crystal strain due to the bending motion of a crystal cantilever. A quantitative analysis shows that good optical sensitivity to the bending motion is obtained if a magnetic field gradient is applied across the crystal during hole burning, and that the resulting opto-mechanical coupling strength is sufficient for observing quantum features such as zero point vibrations.

\end{abstract}


\maketitle


The ability to control and probe physical systems at the quantum mechanical level has been demonstrated in a variety of examples, ranging from single photons and atoms, over currents and voltages in electronic circuits to the motion of mechanical devices. Many of these systems are promising candidates for sensitive and high precision measurements and for transmission, processing and storage of quantum information, but one single system is typically not adequate for all of these functions. This has spurred the interest in so-called hybrid system \cite{hybrid-pnas}  which combine physical components which are separately optimized for different tasks. Outside the technical challenge of handling physically very different systems in a single laboratory experiment, the mismatch of their physical properties (resonance excitation frequencies, spatial overlap, and coherence time scales) presents a main obstacle against the efficient transfer of quantum states between them. For instance, single atoms have microscopic dipole moments and interact in general only weakly with physical observables of mesoscopic quantum systems which occupy orders of magnitude larger spatial volumes. One successful remedy to this weak coupling is to use ensembles of many atomic particles with correspondingly increased coherent coupling strength. This is, indeed, the rationale behind the use of atomic ensembles for optical interfaces and memories and of spin ensembles in conjunction with superconducting circuits~\cite{hybrid-pnas}.
In this article, we address the application of rare-earth ion doped crystals for hybrid quantum technologies. Such crystals have found rich applications in quantum communication protocols where their strong inhomogeneous broadening is favorable for speed and bandwidth.
One particular hybrid technology which holds promise for an efficient coupling between radically different degrees of freedom relies on opto-mechanical interactions~\cite{Aspelmeyer2012,Dantan2014}.  Major achievements such as the ability to prepare a mechanical oscillator in the quantum ground state~\cite{OConnell2010, Teufel2011} have spurred ambitious goals to prepare non-classical  states of motion and use such systems in precision measurements and quantum information applications. While mechanical oscillators can be coupled via light beams to other systems, such as atomic ensembles~\cite{Treutlein2007, Hunger2010} we propose a simpler set-up in which the bending motion of an inorganic crystal is coupled to the optical transitions in an ensemble of rare-earth ion dopants in the same crystal. The crystal strain generated by the harmonic motion of a bulk mechanical oscillation provides an intrinsically stable coupling to the dopant optical properties as, contrary to other coupling mechanisms to externally imposed fields, it circumvents instabilities arising from drifts in position of field source and oscillator. Weak strain coupling to single emitters has been previously observed ~\cite{Yeo2014, Teissier2014, Ovartchaiyapong2014}, but using a rare-earth ion ensemble, we benefit from their strong collective coupling as well as their narrow homogeneous linewidths and long coherence times, allowing unambiguous  observations of quantum features.

We suggest to use spectral hole burning to prepare a transmission window with no optical absorption and obtain the opto-mechanical coupling through the off-resonant, dispersive interaction with ions with transition frequencies outside the hole. As a concrete application, we will focus on $\rm Eu^{3+}$ ions in a $\rm Y_2SiO_5$ host matrix as they exhibit record-narrow optical transitions among solid-state emitters. Despite its long coherence time~\cite{Equall1994}, the 580 nm optical transition $^7F_0 \rightarrow$  $^5D_0$ in $\rm Eu^{3+}$ ions is sensitive to crystal strain~\cite{Thorpe2011}, making it an attractive candidate system for strain-coupled optomechanics. We investigate a crystal in the shape of a micro-meter scale cantilever whose bending produces a local strain-induced perturbation to the ions resonant frequencies.
%
%

\begin{figure}[h]
\centering
\includegraphics[width=80mm]{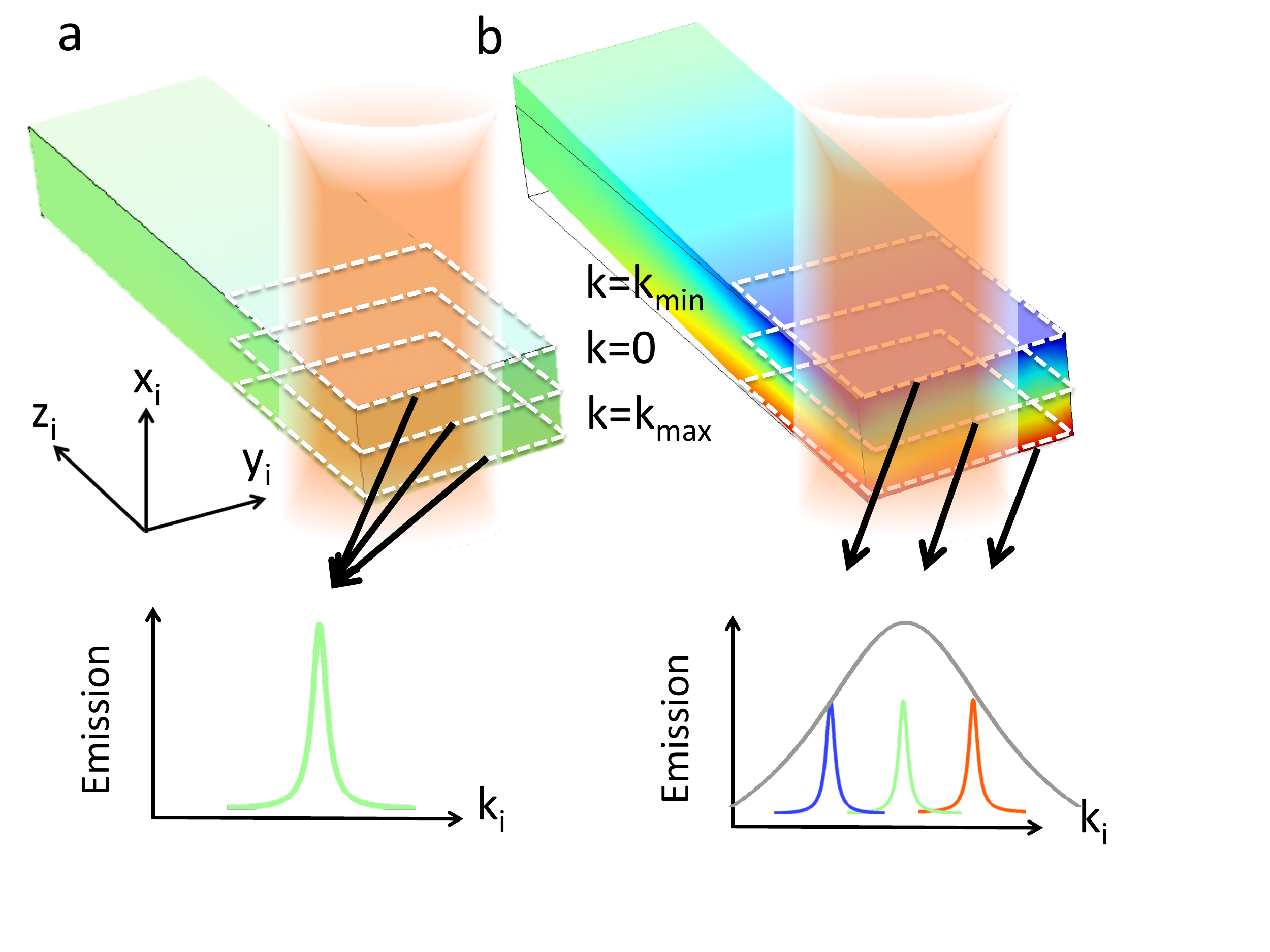}
\caption{\label{peda_fig} Schematics of a cantilever anchored in the foreground of the drawing, and with the far end (a) in its equilibrium position, and (b) bent upwards. The colors on the cantilever indicate different values of the strain: green corresponds to a non-strained material, blue and red to compressive and tensile strain, respectively. The white dashed lines indicate planes of equal strain, neglecting variations along $z_i$. The bottom graphs illustrate how a single emission line is shifted due to the strain (arbitrary units).}
\end{figure}

During bending, there will be a strain gradient across the cantilever, ranging from tensile to compressive strain (see fig.~\ref{peda_fig}), and ions with identical frequencies in the unbent crystal (a) experience different frequency shifts in a bent cantilever (b). In the latter case, ions located in three different layers (subject to identical strain within each layer) experience different strain and the emission lines shift accordingly: if the oscillator is bent upwards, emitters on the top face  experience a compression, whereas emitters on the bottom face are subject to a tensile strain. As the cantilever vibrates, the collective line shape periodically broadens and narrows, and the motion should be readily detected by an optical probe.

However, if static local perturbations of the emitters lead to additional inhomogeneous broadening with a very wide absorption line shape, the strain induced frequency shifts cannot be probed dispersively. To circumvent this, we propose to prepare the ion ensemble by burning of narrow spectral holes with the cantilever at rest, and subsequently use their optical response to the cantilever motion. Spectral holes are formed by resonant optical excitation and decay of the ions into other long-lived (dark) states, typically other hyperfine state within the electronic ground state manifold. Due to the narrow homogenous linewidths, one can scan the excitation laser over a finite frequency interval and burn a spectral hole which must be narrower than the ground state hyperfine splitting to avoid repopulating the hole by subsequent excitation of the dark ions. A weak optical field with a frequency at the center of the hole burnt interval will subsequently interact dispersively with the ions with frequencies outside the spectral hole, and we will in the following propose a method to make this dispersive coupling sensitive to  the cantilever motion.

As indicated by the example in fig.~\ref{peda_fig}, bending of the crystal shifts the transition frequency by an amount proportional to the strain experienced by the individual ions. This implies that, to lowest order, the weak probe field  is not sensitive to the bending, unless we can ensure an asymmetric response of the ion frequency distribution. Such an asymmetry can be prepared by a particular hole burning protocol as discussed in the following sections.

For our analysis we assume a spatially homogeneous ion distribution, and  we also consider the frequency distribution due to the inhomogeneous broadening to be constant within a wide detuning range. We assume that the bending of the beam shifts the resonance frequency of all ions by a quantity proportional to the strain at their position in the crystal.
The bending of the cantilever resonator with thickness $e$ and length $L$ is represented by a single collective coordinate $X$, measured as the vertical displacement of the cantilever tip shown in fig.~\ref{peda_fig}. The ions are all located near the anchoring of the cantilever with spatial coordinates  $(x_i,y_i,z_i)$, where $x_i$ is in the same direction as $X$ and runs from $-e/2$ to $+e/2$, and $z_i$ is along the beam axis, and takes values from $0$ to $L$. From the Euler-Bernouilli theory of beams, we know that the local strain of the matrix around ion $i$ is proportional to both $X$, $x_i$ and $L-z_i$. We optically address only atoms near the base of the cantilever, and we therefore neglect the $z_i$ dependence, so the strain induced frequency shift reads $k x_i X$, with $k$ a constant depending on the beam geometry and physical properties of the doped material.

To enhance the position sensitivity of the cantilever, we prepare our system by applying conventional spectral hole burning while the unbent crystalline cantilever is subject to a magnetic gradient \footnote{Alternatively, one could in principle induce a small, static displacement of the resonator during hole burning, but a magnetic gradient allows for a better control in this particular case. One could also choose to apply an electric field gradient for the same purpose if experimentally more convenient~\cite{Li2016}.}. We  assume a convenient choice of light polarization and magnetic field directions with reference to the crystalline axes so that all ions are shifted by the linear Zeeman effect such that the overall transition frequency of the $i^{th}$ ion at position $(x_i,y_i,z_i)$ can be written as
\begin{equation}\label{freqshift}
\nu_i=\nu_0 + \delta^0_i + k x_i X + g x_i \nabla B=\nu_0+ \delta_i,
\end{equation}
where $\nu_0$ is the chosen central frequency for our hole burnt structure, $\delta^0_i$ is the intrinsic inhomogeneous frequency shift of the $i^{th}$ ions (without the effect of cantilever motion or magnetic field), $\nabla B$ is the externally applied magnetic gradient and $g$ the linear Zeeman effect sensitivity. We finally also assume that $\delta^0_i$ and $x_i$ are uncorrelated.

\begin{figure}
\includegraphics[width=80mm]{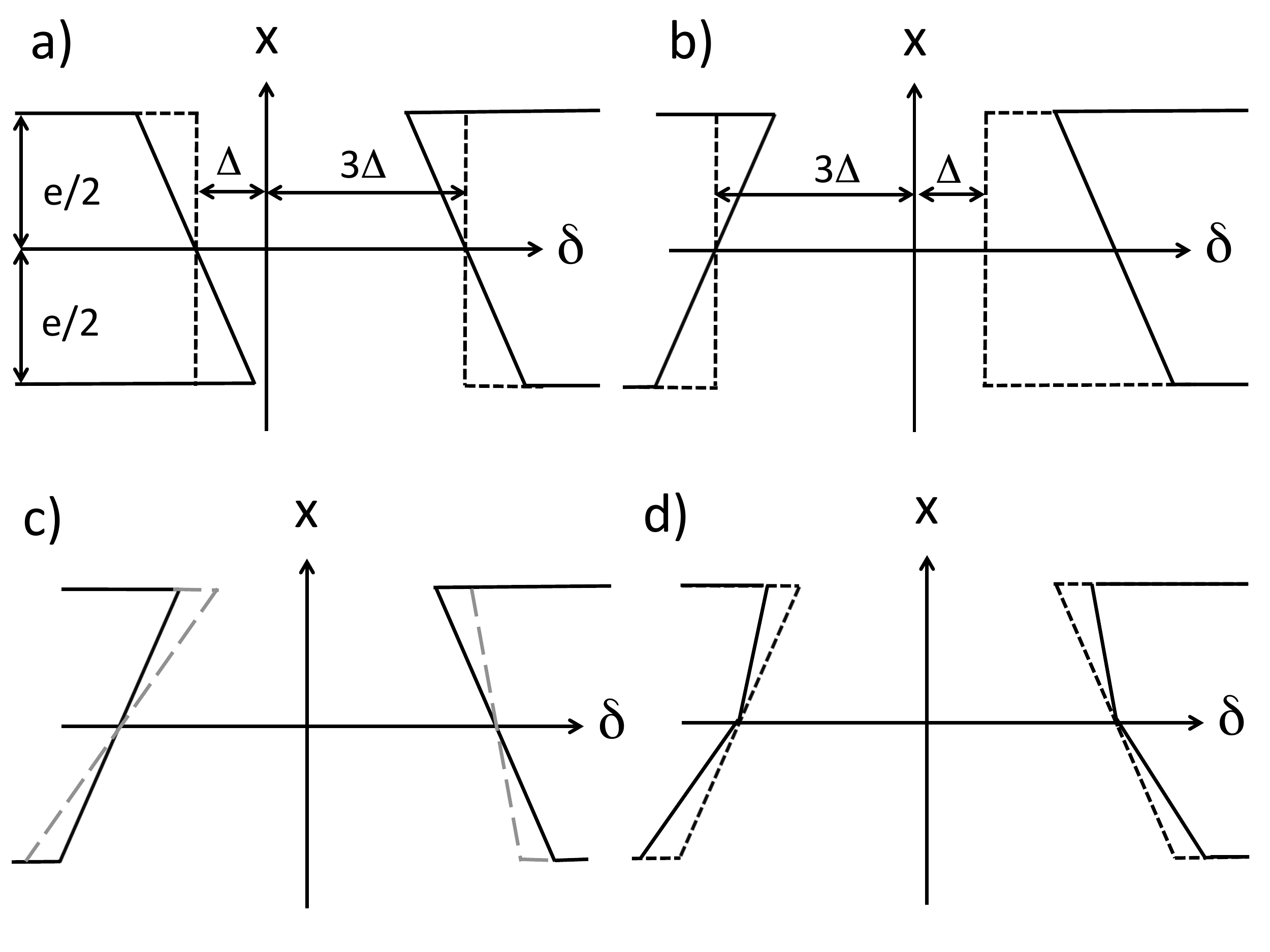}
\caption{\label{holes} {Schematics of the result of the hole burning procedure on the ground state ion distribution as a function of the ion position and "intrinsic" inhomogeneous frequency detuning $\delta_i$. The lines denote the border between the zones with large $|\delta_i|$ values where the ions are left in the ground state, and the central zones where they transferred to the dark state during burning. In the first step of the hole burning procedure (a), the hole is burned in the $[\nu_0-\Delta , \nu_0+3\Delta]$ interval, indicated with dotted lines, while the magnetic gradient $\nabla B_0$ is on. The solid lines show the hole after the magnetic gradient is turned off, realizing the ``right'' edge of the final hole.  The second step (b) realizes the ``left'' edge by hole burning the interval $[\nu_0-3\Delta , \nu_0+\Delta]$ while the magnetic gradient $-\nabla B_0$ is on. The solid lines show the final hole after the 2 steps are completed and the magnetic gradient is turned off. Panel (c) shows the asymmetric effect of bending after burning both edges (in black, the unbent cantilever, and in grey dashed lines, the cantilever bent by a positive amount $X$). Panel (d) shows the effect on the hole due to thermal Brownian motion during the burning procedure: dotted lines correspond to burning at zero temperature and solid lines to burning with finite Brownian motion of the cantilever.}
}
\end{figure}

Our functionalized hole burning procedure consists of two steps:

{\bf 1)}   Apply a static magnetic gradient $\nabla B_0$ and scan the hole burning laser across the spectral range $[\nu_0-\Delta , \nu_0+3\Delta]$ within the inhomogeneous linewidth. The value of $\Delta$ can be chosen arbitrarily, but must fulfill the condition $\Delta\ge g x_i^{\rm max} \nabla B_0$. This step is represented in fig.~\ref{holes}a.

{\bf 2)} Apply the reverse static magnetic gradient $-\nabla B_0$ and scan the hole burning laser across the interval $[\nu_0-3\Delta , \nu_0+\Delta]$. This step is represented in fig.~\ref{holes}b.

As shown in fig.~\ref{holes}, after the magnetic gradient has been switched off, we obtain an inhomogeneous absorption profile of the ions with a transmission hole around the central frequency $\nu_0$, and with a width that depends on the location  $x_i$ in the cantilever but is on average close to $6\Delta$. We are then ready to probe and interact with the system by transmission of a weak laser beam with frequency $\nu_0$, at the center of the hole. In fig.~\ref{holes}c, we show how the frequency shift $k x_i X$ due to bending of the cantilever transforms the shape of the transmission window. In particular, we note that our hole burning procedure ensures a non-vanishing phase shift of the probe beam due to changes in $X$, because contributions from ions with positive $x_i$ are not canceled by the ions with negative $x_i$ as they are further detuned.

At a finite temperature, corrections to the hole shape arise because the thermal Brownian motion of the cantilever amplitude during hole burning exposes the whole inhomogeneous profile to strain distortions like the one shown in fig.~\ref{holes}d. These distortions lead to excitation and decay into dark states of those ions that are brought into resonance with the hole burning laser during their random Brownian motion. We estimate the consequences of this mechanism by assuming that ions with frequencies within $k x_i \bar X$ of the spectral hole in fig.~\ref{holes}a are also transferred to dark states, where $\bar X$ is of same order of magnitude as the root-mean-square thermal excursions of $X$. Fig.~\ref{holes}d shows how the spectral hole is modified by the thermal Brownian motion under hole burning and effectively excludes ions at vertical position $x_i$ that are detuned with respect to $\nu_0$ by an amount $\delta \in [\ell_{L}(x_i),\ell_{R}(x_i)]$, where
$\ell_{L}(x_i)=-3\Delta+ g x_i \nabla B_0 -(+) k x_i \bar X$ and $\ell_{R}(x_i)=3\Delta- g x_i \nabla B_0 +(-) k x_i \bar X$ for positive (negative) values of $x_i$.
		
To describe the interaction between the probe laser field and the hole burnt ensemble, we use a semi-classical approach based on the Maxwell-Bloch equations, which, on the one hand, accounts for the state of the individual atoms by solution of the quantum optical master equation and, on the other hand, describe the damping and dispersion of the Maxwell electric field amplitude due to the interaction with the mean atomic dipole density. Due to the spectral hole, the laser is far off-resonant from all atomic transitions, and we are in the dispersive and perturbative regime of linear excitation of the atoms.

The cantilever is a mechanical resonator of eigenfrequency $\nu_M$, for which we assume a small modulation amplitude. Under this assumption, and imposing the extra condition that $\nu_M^2 \ll |\delta_i|^2$ for adiabatic following (see appendix), the phase shift of the probe field with wavelength $\lambda$ and cross section $A$ interacting with a two-level system with detuning $\delta_i$ and radiative decay rate $\Gamma$ is given by
\begin{equation}\label{phaseshift}
\Delta \varphi_i =  - \frac{\lambda^2\Gamma}{8 \pi A} \frac{1}{\delta_i}
\end{equation}
per atom.

The phase change $\Delta \varphi_i$ of the classical (coherent state) field amplitude is equivalent to a phase change of the same value occurring on each single photon due to the presence of the ion. The combined state of the ion and the stream of photons thus accumulates a quantum phase $\theta$ that changes at a rate $\dot{\theta}= R\Delta \varphi_i$, where $R=\frac{IA}{\hbar \omega_0}$ ($\omega_0=2\pi\nu_0$) is the rate of photons passing through $A$ with intensity $I$. The ion is left in its ground state and the photons leave with unchanged frequency, and the time evolving phase due to the light-matter interaction is equivalent to the light induced AC Stark shift of the ion due to the laserfield,
\begin{equation} \label{int}
V_i = - \hbar \dot{\theta} = - \hbar R\Delta \varphi_i   =  \frac{\sigma_0 I}{\delta_i},
\end{equation}
where $\sigma_0 =\frac{\lambda^3\Gamma}{16 \pi^2 c} $. The expression for $V_i$ has, indeed, the familiar form $I/\delta_i$, valid for large detunings.
We note that we applied a simple two-level description and to evaluate the accurate value of the optical phase shift and atomic energy shift, one must specify the actual transition, the light polarization, and also the index of refraction of the host material.

Since the ion transition frequencies are modified by the crystal strain, their AC Stark shifts become functions of $X$ as well as of the optical intensity, i.e., we obtain the desired optomechanical coupling. The total energy shift of all ions - due to the dispersive light matter interaction, is the  sum of Eq.(\ref{int}) over all ions.

%

To evaluate the dependence of the total interaction energy on the collective cantilever coordinate $X$, we integrate the contributions over the thickness $e$ of the cantilever and the corresponding distribution of $\delta$ from all ions outside the spectral hole:
\begin{multline} \label{xnu_integral}
V(X)=n\sigma_0 I \left[ \int_{-e/2}^{0} dx \left( \int_{-\infty}^{\ell_{L}(x)+k x X}  \frac{d\delta}{\delta}+ \int_{\ell_{R}(x)+k x X}^{\infty}  \frac{d\delta}{\delta}\right)\right.\\
+ \left. \int_{0}^{e/2} dx \left( \int_{-\infty}^{\ell_{L}(x)+k x X}  \frac{d\delta}{\delta}+ \int_{\ell_{R}(x)+k x X}^{\infty}  \frac{d\delta}{\delta}\right)\right].
\end{multline}

Here, $n$ is the density of emitters per unit of frequency and unit vertical distance, and the interaction limits which correspond to the edges of the spectral holes are expressed in terms of the functions defined previously in the text. By applying a first order Taylor expansion of $1/\delta$ in $kxX$, the integrals can be readily evaluated and we obtain

\begin{multline} \label{full_expression}
V(X)=2kXn\sigma_0 I \left[     \frac{g \nabla B_0 e}{(g \nabla B_0)^2-(k \bar X)^2} \right.\\
-  \frac{3\Delta}{(g \nabla B_0+k \bar X)^2} ln \left( 1+\frac{(g \nabla B_0+k \bar X)e}{6\Delta}\right)\\
+ \left. \frac{3\Delta}{(-g \nabla B_0+k \bar X)^2} ln \left( 1+\frac{(-g \nabla B_0+k \bar X)e}{6\Delta}\right)\right].
\end{multline}

Note that for temperatures below 100 mK during hole burning, the Brownian motion is negligible, $k\bar X \ll \Delta$, and the expression for $V(X)$ simplifies, and a third order Taylor expansion in $g\nabla B_0 e/(6\Delta)$ yields
\begin{equation} \label{n}
V(X)=-kXn\sigma_0 I \frac{g\nabla B_0 e^3}{54 \Delta^2}.
\end{equation}

The interaction strength given by this expression is explicitly proportional to $X$ and $I$ as desired for the opto-mechanical coupling. Note that it is also proportional to the bias magnetic gradient $\nabla B_0$, applied during hole burning. The parameter $\nabla B_0$ thus serves as an (adjustable) gain coefficient for the optomechanical coupling. To obtain a maximum coupling strength, it is desirable to match the bias magnetic gradient to the width of the spectral hole $g x_i^{\rm max} \nabla B_0 = \Delta $.

The coupling of the lightfield with the emitters causes an AC Stark shift that depends on the bending of the cantilever and hence results in an displacement $X_{\rm disp}$ of the resonator equilibrium position given by
\begin{equation}\label{staticdisplacement}
X_{\rm disp} = -\frac{dV(X)}{dX}|_{X=0}/m\omega_M^2,
\end{equation}
where $\omega_M=2\pi \nu_M$ is the angular frequency of oscillation of the resonator.

Conversely, the coupling between the light field and the emitters produces a phase shift on the laser. The same proportionality applies between the optical phasehift and interaction energy for the whole system as for a single ion, and we can thus directly write the phase shift of the optical beam as
\begin{equation}\label{sumphaseshift1}
\Delta \varphi = \sum_i \Delta \varphi_i =  \frac{V(X) \omega_0}{IA},
\end{equation}
where the dependence on the intensity $I$ cancels due to the explicit proportionality between $V(X)$ and $I$ in the dispersive coupling regime.

This phase-shift can be detected interferometrically, which provides a readout of the position of the resonator. In particular, for the light intensity that leads to the displacement expressed in Eq. \ref{staticdisplacement}, the corresponding phase shift of the carrier optical frequency is $V(X_{\rm disp}) \omega_0 / IA$.

The vibratory movements, on the other hand,  can be detected on the transmitted laser field by observing spectral sidebands at the mechanical oscillator frequency $\pm\nu_M$ from the carrier. Thermal excitation of the cantilever motion yields a total integrated power of sideband phase fluctuations \footnote  {For a time dependent phase fluctuation, this quantity is the square root of the integrated power spectral density.} equal to $\frac{dV(X)}{dX}|_{X=0} \bar X \omega_0/IA$, which converges to the value $\frac{dV(X)}{dX}|_{X=0} X_{\rm zpf} \omega_0/IA$ at zero temperature. To detect such sidebands requires their amplitude to exceed the shot noise limit for the given optical power and integration time (measurement bandwidth). The resolution of the phase shift is thus ultimately limited by photon absorption and transfer of ions to other long lived states, which gradually  ``overburns'' the spectral hole and destroys it. The transition line in Eu:YSO, however, has a very small natural linewidth $\Gamma$ and therefore, even for high optical power, overburning will not occur before a typical time $\tau=4\pi/\Gamma$, which sets the limit for the integration time. Increasing optical power within the integration time $\tau$ can lead to arbitrarily high resolution. For practical consideration, we choose here to limit the maximal optical power to a few mW, based on the fact that a typical standard commercial close-cycle cryocooler exhibit near 3 K a heat load capacity around 10mK/mW near the sample. Consequently, an optical power above a few mW (part of which may be absorbed during burning or probing phase) may therefore lead to excess heat load and/or thermal instability.  Further improvement of the resolution can be obtained by repeating the experiment (erasing, re-imprinting and probing the spectral hole) and averaging the successive results.


As an example we consider a single-clamped cantilever with the dimensions $100\times10\times10 \mu \rm m^3$ interacting with a laser beam traversing the cantilever near its fixed end for maximum strain as illustrated in fig.~\ref{peda_fig}. We consider a cantilever which consists of $\rm Y_2SiO_5$ (Young Modulus of 135 GP) with an effective mass $m_{\rm eff}=1.1\times 10^{-11}$ kg, and of which the first excited mechanical mode vibrates at $\nu_M$= 890 kHz. The cantilever contains a  0.1 \% doping of $\rm Eu^{3+}$ ions, with a $^7F_0 \rightarrow$  $^5D_0$ transition centered at 580 nm and with $\Gamma= 2\pi\times122$ Hz (and $\Gamma^*$ typically $\le 2\pi \times1$ kHz) at T=3 K. We choose a power of 1 mW and a hole width of $6\Delta$=6 MHz to fulfill the condition for large detuned adiabatic following, see appendix. To calculate the proportionality  constant $k$ we use the values from reference \cite{Thorpe2011}: -211.4 Hz/Pa for crystal site 1, which is the most sensitive of the two non-equivalent sites~\footnote{Note that this value is given for an isotropic pressure, but due to lack of uniaxial strain measurements, we use it here. Measurements on other rare-earth ions in similar host materials reveal comparable isotropic and uniaxial strain sensitivity.}. The sensitivity to magnetic field is $g=3.8$ kHz/G~\cite{Thorpe2013}. In order to maximize the coupling while fulfilling the condition $\Delta \le g \nabla B_0 e$ we choose $\nabla B_0=530$ T/m, which seems possible to achieve in a cryogenic environment over short distances (mm or less) using small superconducting coils.

In this configuration, the static displacement of the tip of the resonator due to the light field amounts to $X_{\rm disp}$=0.4 pm and the corresponding phase shift of the laser (the carrier) equals 0.2 mrad. This shift is easily observable as, for the 1 mW laser power, the shot noise limited phase resolution is 0.45 $\micro$rad within the allowed detection time, before hole-overburning becomes non-negligible (approximately 16 ms for the 122 Hz linewidth). For comparison, a direct reflection of a 1 mW laser on the resonator would give rise to a much smaller static displacement (20 fm).

The amplitude of the Brownian motion at 3~K is $\bar X$=0.2 pm, and the spectral sidebands contain an integrated phase of 0.11 mrad due to this thermal excitation. For an integration time equal to the inverse of the thermal linewidth (25 $\micro$s), the shot-noise limited phase-resolution is 14 $\micro$rad. The thermally excited sidebands are therefore readily observed, even within such short integration time.

Moreover, by increasing the integration time up to the maximum before over-burning, it is possible to observe and measure accurately the detailed shape and size of the sidebands. Zero-point motion of the resonator, averaged over the measurement, induces a small excess integrated phase of 0.4 $\micro$rad in the sidebands. As this value is close to the phase resolution achieved within the maximum integration time before hole-overburning, the shot noise limited resolution is therefore sufficient to observe the effect of the zero-point motion of the mechanical resonator. By using either  dilution fridge or an active laser cooling mechanism (or a combination of these), the temperature can be lowered near a point where the thermal excitation does not dwarf the effect of zero point fluctuations. For example, at 30 mK, the zero point fluctuation induces approximately $10^{-3}$ relative excess integrated phase over the effect of Brownian fluctuations alone. Such a deviation seems measurable, provided sufficient knowledge of the relevant parameters (temperature, Q factor,...). Several measurements at different  temperatures may also be used to estimate the various relevant parameters with the necessary accuracy. Note that the resolution can be further increased by repeating the full hole imprinting and measurement sequence several times, or use optical repumpers to preserve the spectral hole.

A potentially perturbing effect arises due to fluctuations of the laser power. As the static displacement corresponding to 1mW of laser power is approximately 0.4 pm, this laser power must be stable to within $\sim$ 10$^{-4}$  to ensure a perturbation much smaller than the zero point fluctuations ($\sim$ 1 fm), a power stability requirement well within reach of standard stabilization techniques. In addition the laser frequency must also be carefully controlled. The zero-point motion induces a frequency shift of the ions closest to the edge of the resonator of approximately 37 Hz. The probe laser must therefore have a frequency stability substantially better than that, which is well within reach of nowadays commercially available ultra-high finesse Fabry-Perot cavity stabilized lasers (which typically have sub-Hz frequency stability). Note that probing the Brownian motion alone at 3 K exhibits a much less stringent frequency stability requirement at the sub-kHz level.

In summary, we have proposed a hybrid scheme which provides an efficient optomechanical strain based coupling between a mechanical resonator and an ensemble of narrow linewidth ions. The coupling mechanism via a functionalized spectrally burnt hole does not suffer from inhomogeneous broadening of the emitters, nor from the inhomogeneity of the strain across the resonator during displacement. It is even possible to further increase the coupling by exploiting not just one but an ensemble of functionalized spectral holes within the inhomogeneous linewidth, which can be obtained by using a comb of optical frequencies.  The narrow linewidth and dispersive nature of the coupling allows use of a relatively large optical power without destruction of the functionalized hole and observation of the cantilever motion at the quantum level. Moreover, the large back-action of the dispersive coupling on the resonator (more than one order of magnitude superior to that of simple light pressure and comparable to the amplitude of the Brownian motion at 3~K) holds promise for efficient resonator manipulation such as the implementation of active cooling protocols or quantum engineering.

We thank Stefan Kr\"{o}ll, Philippe Goldner and S\'ebastien Bize for discussions, and SS thanks Daniel Est\`eve for help and discussions. KM acknowledges support from the Villum Foundation, YLC from the Ville de Paris Emergence Program and SS from la R\' egion Rh\^one-Alpes (CMIRA Explora-pro). This project has received funding from the European Union's Horizon 2020 research and innovation program under grant agreement No 712721.

\section{Appendix} 
The AC Stark shift  of the atomic ground state level and the phase shift of the transmitted field are both governed by the response of the two-level atomic system to the laser field, and in particular, the mean optical coherence induced in the ions. We denote the lower and upper levels $|g\rangle$ and $|e\rangle$, and we assume that we only excite the ions very weakly, i.e., the ground state population $\rho_{gg} = 1$ to first order in the Rabi frequency $\Omega=DE/\hbar$, where $D$ is the dipole matrix element and $E$ denotes the driving field amplitude. This leads to the equation of motion for the optical coherence
\begin{equation} \label{eq:app1}
\frac{d}{dt}\rho_{eg} = (i\delta_r - \frac{\Gamma}{2})\rho_{eg} + \frac{i\Omega}{2},
\end{equation}
where the decoherence rate $\Gamma$ may include both radiative ($\propto D^2$) and non-radiative contributions, and where we have defined $\delta_r=2\pi\delta$ as the detuning, measured in radians per second.

The steady state solution reads, $\rho_{eg}^{st} = (i\Omega/2)/(\Gamma/2-i\delta_r) \simeq -\Omega/(2\delta_r)$, and the resulting macroscopic polarization of the doped cantilever (with an ion density $n$) is then given by $P=nD\rho_{eg} \simeq -n (\Gamma/\delta_r) E$, leading to the optical phase shift and energy $V \propto P\cdot E$ analyzed in the article.

The purpose of this appendix is to analyze the case where $X$ performs oscillatory motion, $X(t) = X_0 \cos(\omega_M t)$, and the strain induced coupling leads to a temporally modulated detuning, $\delta_r(t) = \delta_r +    \varepsilon \cos(\omega_M t) = \delta_r + \frac{\varepsilon}{2}(e^{i\omega_M t} + e^{-i\omega_M t})$. Due to the long lifetime of the ion coherence and excited state, we cannot merely assume that the coherence $\rho_{eg}$ attains the stationary solution $\rho_{eg}^{st}$, evaluated at the time dependent value of the detuning. We can assume, however, that the ions reach a periodic steady state, and for weak modulation, we make the Ansatz,
\begin{equation}
\rho^{pss}_{eg}(t) = a^0 + a^+ e^{i\omega_M t} + a^- e^{-i \omega_M t}.
\end{equation}
Applying this Ansatz in \eqref{eq:app1}, and isolating terms $\propto e^{\pm i\omega_M t}$ leads to the relations, $a^+= \frac{\varepsilon/2}{\omega_M - \overline{\delta}}a^0$, $a^-= \frac{-\varepsilon/2}{\omega_M + \overline{\delta}}a^0$ where $\overline{\delta}=\delta_r+i\Gamma/2$. Inserting these results in the equation for the non-oscilating terms in the periodic solutions to \eqref{eq:app1}, we obtain, $a^0 \simeq -\Omega/(2\delta_r)$, which holds to first order in the small modulation amplitude of the detuning $\varepsilon$.

Collecting terms, the periodically varying coherence writes,
\begin{equation} \label{eq:coherence}
\rho_{eg}^{pss}(t) = \frac{-\Omega}{2\delta_r} \bigg[1+ \frac{2 \varepsilon \overline{\delta} \cos(\omega_M t)}{\omega_M^2 - \overline{\delta}^2} +
\frac{2 i\varepsilon \omega_M \sin(\omega_M t)}{\omega_M^2 - \overline{\delta}^2}\bigg].
\end{equation}

If $\omega_M^2 \ll |\overline{\delta}|^2$, and if $\Gamma \ll |\delta_r|$, the expression simplifies,
\begin{equation} \label{eq:coherencead}
\rho_{eg}^{pss}(t) \simeq  \frac{-\Omega}{2\delta_r } \bigg[1 - \frac{2\delta_r \varepsilon \cos(\omega_M t)}{\delta_r^2}\bigg]
\simeq \frac{-\Omega} {2\delta_r(t)},
\end{equation}
and the coherence, indeed, follows the stationary solution adiabatically in this coupling regime.

When $\omega_M$ is comparable to or larger than $|\delta_r|$, we see a more complicated dependence of the real (dispersive) and imaginary (absorptive) part of the coherence, and also a weaker dependence on the position in the high frequency limit, where the coherence is unable to follow the time dependent steady state.

\end{document}